\documentclass[12pt]{article}
\usepackage{epsf, cite, amssymb}
\usepackage{epsfig}

\makeatletter
\renewcommand\section{\@startsection {section}{1}{\z@}%
                                   {-3.5ex \@plus -1ex \@minus -.2ex}
                                   {2.3ex \@plus.2ex}%
                                   {\normalfont\large\bfseries}}
\renewcommand\subsection{\@startsection{subsection}{2}{\z@}%
                                     {-3.25ex\@plus -1ex \@minus -.2ex}%
                                     {1.5ex \@plus .2ex}%
                                     {\normalfont\bfseries}}
\makeatother

\let\non\nonumber

\newcommand{\bea}{\begin{eqnarray}}
\newcommand{\eea}{\end{eqnarray}}
\newcommand{\be}{\begin{equation}}
\newcommand{\ee}{\end{equation}}

\newcommand{\hlf}{\frac{1}{2}}

\newcommand{\mO}{\Omega}

\newcommand{\G}{\Gamma}

\newcommand{\e}{\epsilon}

\newcommand{\dd}{\delta}

\newcommand{\rr}{\rightarrow}
\newcommand{\m}{\mu}

\newcommand{\p}{\partial}

\newcommand{\tr}{{\rm Tr}}

\newcommand{\C}[1]{$(\ref{#1})$}

\def\IZ{\relax\ifmmode\mathchoice
{\hbox{\cmss Z\kern-.4em Z}}{\hbox{\cmss Z\kern-.4em Z}}
{\lower.9pt\hbox{\cmsss Z\kern-.4em Z}} {\lower1.2pt\hbox{\cmsss
Z\kern-.4em Z}}\else{\cmss Z\kern-.4em Z}\fi}
\def\IR{\relax{\rm I\kern-.18em R}}

\def\one{{\hbox{ 1\kern-.8mm l}}}

\def\tr{{\rm tr\,}}

\newlength{\bredde}
\def\slash#1{\settowidth{\bredde}{$#1$}\ifmmode\,\raisebox{.15ex}{/}
\hspace*{-\bredde} #1\else$\,\raisebox{.15ex}{/}\hspace*{-\bredde}
#1$\fi}

\newsavebox{\zzzbar}
\sbox{\zzzbar}
  {\setlength{\unitlength}{0.9em}
  \begin{picture}(0.6,0.7)
  \thinlines
  \put(0,0){\line(1,0){0.6}}
  \put(0,0.75){\line(1,0){0.575}}
  \multiput(0,0)(0.0125,0.025){30}{\rule{0.3pt}{0.3pt}}
  \multiput(0.2,0)(0.0125,0.025){30}{\rule{0.3pt}{0.3pt}}
  \put(0,0.75){\line(0,-1){0.15}}
  \put(0.015,0.75){\line(0,-1){0.1}}
  \put(0.03,0.75){\line(0,-1){0.075}}
  \put(0.045,0.75){\line(0,-1){0.05}}
  \put(0.05,0.75){\line(0,-1){0.025}}
  \put(0.6,0){\line(0,1){0.15}}
  \put(0.585,0){\line(0,1){0.1}}
  \put(0.57,0){\line(0,1){0.075}}
  \put(0.555,0){\line(0,1){0.05}}
  \put(0.55,0){\line(0,1){0.025}}
  \end{picture}}

\newcommand{\ena}{\end{eqnarray}}
\newcommand{\beqa}{\begin{eqnarray}}
\newcommand{\eeqa}{\end{eqnarray}}

\newcommand{\half}{\frac{1}{2}}
\newcommand{\eq}[1]{(\ref{#1})}

\def\G{\Gamma}

\def\e{\epsilon}

\def\m{\mu}

\def\G{\Gamma}

\begin{document}
\begin{titlepage}

\begin{center}

{June 28, 2005}
\hfill                  hep-th/0506180

\hfill ITFA-2005-21, EFI-05-04

\vskip 2 cm
{\Large \bf A Matrix Big Bang}\\
\vskip 1.25 cm { Ben Craps$^{\,a}$\footnote{email address:
 bcraps@science.uva.nl},  Savdeep Sethi$^{\,b}$\footnote{email address:
 sethi@theory.uchicago.edu} and Erik Verlinde$^{\,a}$\footnote{email address:
 erikv@science.uva.nl}
}\\
{\vskip 0.5cm $^{a}$ Instituut voor Theoretische
Fysica, Universiteit van Amsterdam,
Valckenierstraat 65,
1018 XE Amsterdam,
The Netherlands \\
\vskip 0.5cm
$^{b}$ Enrico Fermi Institute, University of Chicago,
Chicago, IL
60637, USA\\}

\end{center}

\vskip 2 cm

\begin{abstract}
\baselineskip=18pt

The light-like linear dilaton background represents a particularly
simple time-dependent $1/2$ BPS solution of critical type IIA superstring
theory in ten dimensions. Its lift to M-theory, as well as its Einstein
frame metric, are singular in the sense that the geometry is geodesically
incomplete and the Riemann tensor diverges along a light-like subspace of
codimension one. We study this background as a model
for a big bang type singularity in string theory/M-theory. We construct
the  dual Matrix theory description in terms of a (1+1)-d supersymmetric
Yang-Mills theory on a time-dependent world-sheet given by the Milne
orbifold of (1+1)-d Minkowski space. Our model provides a framework in which
the physics of the singularity appears to be under control.

\end{abstract}

\end{titlepage}

\pagestyle{plain}
\baselineskip=19pt
\section{Introduction}

One of the outstanding questions facing string theory is how to
describe a cosmological singularity like the big bang. For recent
cosmological scenarios where the big bang singularity plays a crucial role,
see for instance~\cite{cosmo}. What prior
work has taught us is that perturbative string theory breaks down on
many toy model space-times that include space-like and light-like curvature
singularities \cite{
Liu:2002ft, Liu:2002kb, Lawrence:2002aj, Horowitz:2002mw, Berkooz:2002je}.  See~\cite{related, Nekrasov:2002kf, Elitzur:2002rt,
 Berkooz:2004re} for some related work.
To capture the physics of the  singularity, a complete
non-perturbative description of string theory appears to be necessary.%
\footnote{See,
however, the very recent paper \cite{McGreevy:2005ci}, which claims that certain
space-like singularities are replaced by a tachyon condensate phase within perturbative
string theory.}
In
this work, we will present a particularly clean example of a
cosmological singularity that admits a holographic dual description
via Matrix Theory~\cite{Banks:1996vh} (for reviews, see for instance
\cite{matrixrev}). Prior examples of holographic descriptions, in the sense of AdS/CFT, appear
in~\cite{Elitzur:2002rt, holo}. See \cite{matrixcosmology} for some other
ideas relating Matrix theory and cosmology.

The backgrounds that we will consider are linear dilaton backgrounds
which are key ingredients in some of the oldest exact solutions
of string theory~\cite{LD}. The dilaton,  $\phi$, is identified with a
direction in space-time. If the direction is time-like, the solution
is cosmological and non-supersymmetric. By
definition, we lose perturbative control over these backgrounds when the string
coupling
\be g_s = e^\phi \ee
becomes large.

In this work, we want to consider a simple variant of these
cosmologies where we choose light-cone coordinates in space-time and identify,
\be \label{lindil}
\phi = - Q X^{+},
\ee
where $Q$ is a constant.
This kind of dilaton profile appears as an ingredient in many supergravity
solutions like some plane-wave backgrounds.
To construct a solution of string theory, we also need to specify a
$10$-dimensional space-time metric. This metric could describe some non-trivial
compactification. For simplicity, we will take flat Minkowski space with
coordinates $X^\m = (X^+, X^-, X^i)$ and metric,
\be ds_{10}^2 = -2 dX^+ dX^- + \sum_{i} (dX^i)^2 \ee
as our $10$-dimensional string metric.

This background is a remarkably
simple, time-dependent string solution. Note that the parameter $Q$
appearing in \C{lindil}\ can be scaled to any non-zero value using the boost
symmetry
$$ X^+ \rr \alpha X^+, \qquad  X^- \rr \alpha^{-1} X^-, $$
where $\alpha$ is non-zero.
To see that flat space is
still a string solution in the presence of this linear dilaton, we
only need to note that a light-like linear dilaton (unlike a space-like or
time-like linear dilaton) makes no
contribution to the conformal anomaly.

{}From the perspective of string frame, the only time-dependence appears in the coupling constant.
However, the corresponding Einstein
frame metric is given by
\be\label{einstein}
ds_{E}^2 = e^{Q X^+/2} \, ds_{10}^2.
\ee
Viewed in Einstein frame, this space-time originates at a big bang as $X^+ \rr -\infty$ since the
scale factor goes to zero. In the following section, we will study this solution as a model for a big bang, and
describe perturbative string quantization in this background.
In section~\ref{matrix}, we derive the Matrix description of this background which involves Matrix strings
propagating on a time-dependent world-sheet. The world-sheet is described by the two-dimensional metric
\be\label{confmet}
ds^2 = e^{2\eta} \left( -d\eta^2 + dx^2 \right)
\ee
with $x\sim x+2\pi$. This metric describes the future quadrant of the Milne orbifold, which can be thought of as
flat space with a boost identification.
The curvature singularity of the metric~\C{einstein}\ corresponds via Matrix theory to the Milne singularity
at $\eta=-\infty$, where the $x$-circle shrinks to zero size. This is the ``big bang''. Time evolution from
the big bang to the asymptotic regime corresponds in the Matrix description to renormalization group flow from
the UV to the IR. The physics near the big bang is described by weakly-coupled Yang-Mills theory. In
section~\ref{decargument}, we argue that our Matrix description remains decoupled from gravity even at
the singularity.
At late times,
the matrix degrees of freedom re-organize themselves into weakly-coupled strings, and a
conventional space-time picture emerges.

It is worth noting that the Milne orbifold has been studied as a space-time background for closed string
propagation. What can be concluded from this work is that perturbative string theory breaks down because
of large gravitational backreaction from the singularity. In our case, the Milne orbifold appears as the
Matrix string world-sheet. Matrix string theory should capture the non-perturbative physics of the
space-time singularity.

 In the final section, we mention a
few of the possible generalizations of the light-like linear dilaton background.

\section{The Light-like Linear Dilaton}

\subsection{The light-like linear dilaton as a big bang cosmology}

We begin by considering the light-like linear dilaton as a background of type
IIA string
theory. This defines an exact CFT that describes string
propagation in flat space-time with a varying string coupling
given by
\be\label{ggssn}
g_s=e^{-QX^+}.
\ee
The space-time theory is free at late times ($X^{+}\rightarrow\infty$), and strongly coupled
at early times ($X^{+}\rightarrow -\infty$).

This background preserves one-half of the $32$ flat space
supersymmetries. To see this, we need to check the supersymmetry
variations of the gravitino and dilatino. Only the dilatino,
$\lambda$,  feels
the presence of this linear dilaton background via the term in its
supersymmetry variation,
\be
\dd \lambda \sim \G^+ \p_+ \phi\, \e,
\ee
where $\e$ is the supersymmetry parameter. However, there are $16$
solutions to the condition
\be \G^+ \e =0, \ee
so half the supersymmetry is preserved.
This is rather crucial since as $X^+\rr -\infty$, we  have
enough control from supersymmetry to determine a good strong coupling
description. The spectrum in the weak coupling regime where $X^+\rr
\infty$ is determined from the perturbative string quantization to be described in
section~\ref{pertstring}.

As $g_s$ becomes large, we expect
this background to lift to a solution of M-theory. The
$11$-dimensional metric
\be\label{11met}
ds^2 = e^{ 2 Q X^+/3 } ds_{10}^2 + e^{-4 Q X^+/3} (dY)^2,
\ee
with $Y$ the eleventh direction, governs the strong coupling limit of
this background. We should check
that the background defined by \C{11met}\ is not trivial. We define
an orthonormal basis of 1-forms
\be\label{noncoord}
e^i = e^{Q X^+/3} dX^i, \qquad e^+ = e^{QX^+/3} dX^+, \qquad e^- =
e^{QX^+/3} dX^-, \qquad e^{y} = e^{-2QX^+/3} dY \ee
with respect to which the metric takes the canonical form
\be ds_{11}^2
= -2 e^+ e^- + (e^i)^2 + (e^{y})^2.
\ee
Up to symmetry, the corresponding spin connection has non-vanishing components
\be
\omega_{i +} =  {Q\over 3} e^{-QX^+/3} \, e^i, \qquad \omega_{y +} =
     - {2Q\over 3} e^{-QX^+/3} \, e^{y}, \qquad \omega_{-+}=- {Q\over 3}
      e^{-QX^+/3} e^+. \ee
The non-vanishing curvature $2$-forms are
\bea
R_{+i} &=& {Q^2\over 9} e^{-2QX^+/3} e^+ \wedge e^i
\nonumber
, \cr
R_{y+} &=& {8  Q^2 \over 9} e^{-2QX^+/3} e^+ \wedge e^y,
\eea
with respect to the orthonormal basis \C{noncoord}.
The non-vanishing components of the Riemann tensor in a
coordinate basis are (up to symmetry)
\bea
R_{+i+i} &=& {Q^2\over 9} e^{2QX^+/3}, \cr
R_{+y+y} &=&- {8  Q^2 \over 9} e^{-4QX^+/3}.
\eea
It is easy to see from these equations that the Ricci tensor vanishes, as it
should for a purely gravitational M-theory solution.

It might appear that the $11$-dimensional metric~\C{11met}\ has a `singularity'
at both $X^+\to+\infty$ and
$X^+\to -\infty$ since in both limits some metric components go to zero.
The difference between these two limits is, however, that the
$X^+\to-\infty$ singularity occurs at finite geodesic distance, while the
$X^+\to\infty$ singularity is at infinite distance.

The presence of the finite distance $X^+\to-\infty$ singularity implies  that
the space-time is geodesically incomplete. Namely, some geodesics terminate
at finite affine parameter. This is most easily seen for the lines $X^-= {\rm
const.}$, $X^i={\rm const.}$, which are geodesics. The geodesic equation
in this case is
\be
{d^2X^+\over d\lambda^2}+\Gamma^+_{++} \left({dX^+\over d\lambda}\right)^2=0,
\ee
where $\Gamma^+_{++} = 2Q/3$. This can be integrated to give
\be
e^{{2\over 3}QX^+}\left({dX^+\over d\lambda} \right)= {\rm const.}
\ee
and hence the affine parameter is (up to an affine transformation)
\be
\lambda = e^{{2\over 3}QX^+}.
\ee
We thus find that the point $X^+\to-\infty$ corresponds to $\lambda=0$, and
hence it has finite affine distance to all points in the interior. Note
that the other `singularity'
at $X^+\to \infty$ is indeed at infinite affine parameter, $\lambda=\infty$, so
it represents an asymptotic region in which the eleventh dimension
happens to curl up to zero size.

One can write the metric in terms of the affine parameter $\lambda$ for
$\lambda>0$ as
\be\label{11metbis}
ds^2 =- {3\over Q} d\lambda dX^- + \lambda ds^2_8 + \lambda^{-2} dy^2.
\ee
In terms of these coordinates, the non-vanishing components of the
Riemann tensor are
\bea
R_{\lambda i\lambda i} &=& {1\over 4\lambda}, \nonumber \\
R_{\lambda y\lambda y} &=&- {2\over\lambda^4}, \label{11dRiemann}
\eea
which clearly shows that there is a curvature singularity at
$\lambda=0$, where an inertial observer experiences divergent
tidal forces.

It does not make
sense to consider the metric \eq{11metbis} for $\lambda<0$ because the signature
of the eight transverse dimensions changes sign. To extend to
$\lambda<0$ in a sensible way, one might try replacing the
$\lambda$ in front of $ds_8^2$ by its absolute value $|\lambda|$.
However, this extension is ad hoc without some additional input beyond general
relativity about how to treat the curvature singularity. What we can conclude is that
there is truly a singularity in the classical gravity description of the light-like linear
dilaton background.

In fact, this same conclusion also applies to the 10-d
desciption in Einstein frame: namely, there exists a singularity at finite geodesic distance.
To see this, rewrite the Einstein metric \eq{einstein} in terms of
its affine parameter
\be
u=e^{\half QX^+}
\ee
and a new coordinate $v=X^-$:
\be
ds^2_E=-{4\over Q} du\,dv+u\sum_i(dX^i)^2.
\ee
Defining an orthonormal basis
\be
e^i=u^{1/2} dX^i ,\ \ \ \ e^u={2\over Q}du,\ \ \ \ e^v=dv,
\ee
we find the following non-vanishing components of the spin
connection:
\be
\omega^i{}_u={Q\over 4u} e^i,
\ee
and of the curvature two form:
\be
R^i{}_u={Q^2\over 16u^{2}}\,e^i\wedge e^u.
\ee
In a coordinate basis, we find
\be
R_{iuiu}={1\over 4u},
\ee
which is indeed singular at $u=0$.

In Einstein frame, the Ricci tensor is non-zero:
\be\label{ricci}
R_{uu}={2\over u^2}.
\ee
This non-vanishing Ricci tensor is supported by the dilaton, which, unlike in the string frame,
has a non-zero stress-energy tensor and thus contributes to Einstein's equations. We have
\be
\phi=-QX^+=-2 \log u,
\ee
and thus
\be
T_{uu}={1\over 2}(\partial_u\phi)^2= {2\over u^2}.
\ee
This can be interpreted as follows: the singular nature of the $R_{\lambda y\lambda
y}$ component of the 11-d Riemann tensor \eq{11dRiemann} is transferred to  the stress-energy
tensor of the dilaton, and hence by Einstein's equations to the Ricci
tensor \eq{ricci}.

\subsection{Perturbative string theory}
\label{pertstring}
We now describe some of the properties of the light-like
linear dilaton solution in perturbative string theory.
We start with the bosonic string in $D$ dimensions. The world-sheet fields
$X^\mu$ are free, but the time-dependence of the
dilaton is reflected in a modified world-sheet stress tensor,
\be\label{modstress}
T(z)=-\partial X_i
\partial X^i+2\partial X^+
\partial X^- -Q\partial^2 X^+.
\ee
The central charge is unmodified,
\be\label{central}
c=D,
\ee
and $Q$ is a free parameter.

Since the world-sheet theory is free and the string coupling is
small at late times, it is not difficult to construct the physical states.
As in flat space-time with constant dilaton, states are labeled
by momentum $p_\mu$ and an oscillator contribution. The corresponding
vertex operators have the form
\be\label{vertope}
V=e^{ip_\mu X^\mu}
P_N(\partial X^\mu,\bar\partial X^\mu,\cdots),
\ee
where $P_N$ is a polynomial in derivatives of the world-sheet
fields $X^\mu$ of total (left and right) scaling dimension $N$. We take the zero mode part of the
vertex operator to be a plane
wave. Physical states correspond to Virasoro primaries
of the form \C{vertope}\ with scaling dimension one.
{}From \C{modstress}, it follows that the scaling dimension of $V$ is
\be\label{lzero}
L_0={1\over 4}p_i^2-{1\over2}p^+(p^- +iQ)+N.
\ee
The non-standard
contribution of $p^+$ to the scaling dimension is easy to
understand. In string theory, the zero mode part of
vertex operators for the emission of string modes have the general form
\be\label{verwave}
V=g_s\Psi,
\ee
where $\Psi$ is the wavefunction of the state.
Usually, the factor of $g_s$ in (\ref{verwave})
can be neglected since it is constant, but here according to \C{ggssn}\ it is
time-dependent and therefore needs to be retained.
The vertex operator (\ref{vertope}) corresponds to the wavefunction
\be\label{psiwave}
\Psi(\vec X, X^+,X^-)=e^{i\vec p\cdot \vec X-ip^+X^-
-i(p^- +iQ)X^+}.
\ee
Thus, the light-cone energy is
\be\label{zeromom}
E^-=p^- +iQ
\ee
and \C{lzero}\ reads
\be\label{lzeromodif}
L_0={1\over 4}\left(\vec p^2-2p^+E^-\right)+N.
\ee
The mass shell condition then becomes
\be\label{masss}
m_{\rm eff}^2\equiv 2p^+E^--\vec p^2=4(N-1).
\ee
The classical evolution of fields in this background is easy to describe. For example, consider a scalar
{}field $T$ with mass
$m$ in the light-like linear dilaton background.
The Lagrangian is proportional to
\be\label{lscalarn}
{\cal L}=\hlf e^{2QX^+} (2\partial_+T\partial_-T-\partial_i T\partial_i
T-m^2 T^2).
\ee
The equation of motion of $T$ is
\be\label{eomn}
(2\partial_+\partial_--\partial_i\partial_i+2Q\partial_-+m^2) T=0.
\ee
A basis of solutions is given by
\be\label{basissoln}
T(X^+,X^-,\vec X)=e^{-QX^+}e^{-ip^+X^--iE^-X^++i\vec k\cdot\vec X},
\ee
with
\be\label{eeeen}
-2p^+E^-+\vec p^2+m^2=0.
\ee


\subsection{Light-cone string field theory}

The calculation of the perturbative string amplitudes of the light-like
linear dilaton background becomes particularly simple in the light-cone
gauge. Of course, given that the dilaton itself picks a preferred
light-cone direction, one does not even break Lorentz invariance by
making the usual gauge choice $X^+=p^+\tau$ on the world-sheet. The world-sheet
theory for the transverse coordinates is completely identical to that of
flat space superstring theory.

As is well known from the old literature, one can represent a perturbative
string amplitude in terms of a sum over light-cone diagrams. The
contribution of each diagram is expressed as an integral over the
positions $\tau_i$  of the joining and splitting operators on the
world-sheet. For a genus $g$ contribution to a $n$-string scattering
amplitude the number of these vertex operators is $2g-2+n$.  The effect of
the linear dilaton is that the coupling constants now becomes a function
of the light-cone coordinate $\tau$ on the world-sheet. Specifically, every
joining/splitting operator gets multiplied by $e^{Qp^+\tau_i}$. Hence
the overall amplitude, before integrating over the $\tau_i$, gets
multiplied by
\be
\prod_{i=1}^{2g-2+n}e^{-Qp^+\tau_i}\equiv e^{-(2g-2+n)Qp^+\tau_{*}}.
\ee
Here $\tau_{*}$ is the average of the insertion points $\tau_i$.
Because the world-sheet theory is translation invariant in $\tau$, the rest
of the integrand only depends on the relative differences of the positions
 $\tau_i$ of the joining/splitting vertices. This fact can be exploited by
separating the integration over the $\tau_i$ into the integral over the
relative positions multiplied by the integral over $\tau_{*}$. The integral over
the relative positions precisely gives the usual amplitude in flat space.
We thus obtain the following simple relation between the string amplitudes
in the light-like linear dilaton background and the corresponding flat space
amplitudes:
\be
A^{g,n} = A^{g,n}_{\rm flat} \int_{-\infty}^{+\infty} d\tau_{*}
e^{-(2g-2+n)Qp^+\tau_{*}}.
\ee
Clearly, the integral diverges even before summing over the genus.
Hence, one clearly has to introduce a cut-off in the $\tau^*$ integral,
keeping it away from $\tau=-\infty$. But even when we take
$\tau_{*}>\tau_c$ one finds that the effective coupling $g_s^{eff}\sim
e^{-Qp^+\tau_c}$ can become large when $\tau_c$ is negative. Therefore
another description is required in this region. We will provide such a description
in the next section.

\section{Matrix String Description}
\label{matrix}

We will begin our discussion of Matrix theory by taking the flat space Matrix
string action and inserting the time-dependent string coupling, $g_s = \exp(-QX^+)$. This leads
directly to supersymmetric Yang-Mills on the Milne orbifold as the Matrix description of the light-like
linear dilaton space-time. In section~\ref{derivation}, we will
provide an independent derivation leading to this same conclusion. This derivation will allow
us to describe the regime of validity of the Matrix description.

Matrix string theory is described by a
$(1+1)$-d super-Yang-Mills (SYM) theory with 16 supercharges. The
action follows from dimensional reduction of $(9+1)$-d SYM theory.
It contains eight matrix-valued fields $X^i$ representing the
transverse bosonic coordinates, as well as eight matrix-valued
spinor coordinates $\Theta^a$.  The action is
\cite{Motl:1997th, Banks:1996my, Dijkgraaf:1997vv}
\be \label{sym} S = {1\over 2\pi \ell_s^2}
\int {\tr}\left( {1\over 2}(D_\mu X^i)^2 + \theta^T
{D\!\!\!\!\slash{}} \,\theta + { g_s^2 \ell_s^4 \pi^2} F_{\mu\nu}^2 - {1\over
4 \pi^2 g_s^2\ell_s^4}[X^i,X^j]^2 + {1\over 2\pi g_s\ell_s^2} \theta^T\gamma_i
[X^i,\theta]\right).
\ee
The metric on the world-sheet is flat,
{ i.e.}\ $\!\eta_{\mu\nu}=\mbox{diag}(-1,1)$, and the spatial
coordinate $\sigma$ on the world-sheet has a fixed periodicity
equal to $2\pi\ell_s$.
 Notice that the Yang-Mills coupling constant, which is dimensionful in
$(1+1)$ dimensions, is here identified with
 the inverse product of the string
 length and the string coupling,
\be \label{ymcoupling} g_{YM}\equiv {1\over g_s\ell_s}. \ee
In the IR limit the
SYM theory become strongly coupled and, as shown in~\cite{Dijkgraaf:1997vv},
reduces to the perturbative description of the type IIA superstring. In
the UV, however, the SYM theory is weakly coupled.

In the light-cone gauge the world-sheet time coordinate is proportional to
the space-time null coordinate $X^+$.  We can thus describe the light-like linear dilaton
background in a simple way
by allowing the string coupling to depend on the world-sheet time $\tau$ via a relation like
$g_s=e^{-Q\tau}.$ In section~\ref{derivation}, we will determine the precise
proportionality constant between $X^+$ and $\tau$ leading to~\C{gym}.

It thus appears that we are dealing with a SYM theory with a time-dependent
coupling constant. However, there is another way to view
this result; namely, by changing the geometry on the world-sheet.
Unlike the usual string action,  the matrix string action is not
conformally invariant: rescaling the metric by a function
$f(\tau)^2$ changes the terms involving the coupling $g_s$ in such
a way that $g_s$ gets multiplied by $f(\tau)^{-1}$. Thus, we
conclude that the matrix string description of the light-like
linear dilaton background is given by $(1+1)$-d SYM theory with
fixed coupling, but on a world sheet with geometry
\be
ds^2 =e^{2Q\tau}(-d\tau^2+d\sigma^2).
\ee
In fact, this metric is flat since it reduces to the usual Minkowski
metric $ ds^2=-2 d\xi^+d\xi^-$ through the substitution
\be
\xi^\pm  = {1\over \sqrt{2} Q}e^{Q(\tau\pm\sigma)}.
\ee
However, since the coordinate $\sigma$ is periodic modulo $2\pi\ell_s$,
the $(1+1)$-d Minkowski space described by the coordinates $(\xi^+,
\xi^-)$ turns into the Milne orbifold because of the
identifications
\be
\xi^\pm \equiv e^{\pm 2\pi Q\ell_s}\xi^\pm.
\ee

\subsection{A more detailed derivation}
\label{derivation}

We would like to extend the derivation
of Matrix theory given in~\cite{Seiberg:1997ad} (see also~\cite{Sen:1997we}) to our time-dependent example.
We start with the ten-dimensional string metric
\be\label{Mink}
ds^2 = -2 dX^+ dX^- + \sum_{i=1}^8 (dX^i)^2
\ee
and the light-like linear dilaton
\be\label{lld}
\phi = -QX^+.
\ee
In discrete light-cone quantization (DLCQ), we make the identification
\be\label{DLCQ}
X^-\sim X^- + R
\ee
and focus on a sector with $N$ units of light-cone momentum,
\be
p^+={2\pi N\over R}.
\ee
In~\cite{Seiberg:1997ad}, the theory with the identification~\C{DLCQ}\ was defined as
a limit of a space-like compactification, where the shift~\C{DLCQ}\ of $X^-$ is
accompanied by a small shift of $X^+$. However, shifting $X^+$ is not a symmetry
of our background~\C{lld}, so we have to define the DLCQ in a different way. To
that effect, we single out one direction $X^1$ from among the $X^i$ and make the
identification
\be\label{identify}
(X^+,X^-,X^1) \sim (X^+,X^-,X^1) + (0,R,\epsilon R),
\ee
where in the end we will take $\epsilon\to0$.
The Lorentz transformation
\bea
X^+ &=& \epsilon x^+,\nonumber\\
X^- &=& {x^+ \over 2\epsilon}  + {x^- \over \epsilon} + {x^1 \over
\epsilon},\nonumber\\
X^1&=& x^+ + x^1\label{basistransf}
\eea
puts the background in the form
\bea
ds^2& =& -2 dx^+ dx^- + \sum_{i=1}^8 (dx^i)^2,\\
\phi &=& -Q\epsilon x^+,
\eea
with the identification
\be\label{identifybis}
x^1 \sim x^1 + \epsilon R.
\ee
In this background, we focus on a sector with $N$ units of momentum in the $x^1$
direction. After a T and an S duality, and introducing
\be
r \equiv {\epsilon R \over 2\pi\ell_s},
\ee
we are studying a sector with $N$ D1-branes wrapped around $x^1$ in the
type IIB background
\bea
ds^2& =& r e^{\epsilon Q x^+} \left\{ -2 dx^+ dx^- + \sum_{i=1}^8 (dx^i)^2\right\},\label{metricD1}\\
\phi &=& \epsilon Q x^+ + \log r\label{dilneg},
\eea
with the identification
\be
x^1 \sim x^1 + {2\pi\ell_s \over r}.
\ee
This is now a theory of D1-branes in a background where the string coupling becomes weak near the
big-bang, and strong at late times. It is worth stressing that this is opposite to the behavior of the string
coupling in our original background~(\ref{Mink}, \ref{lld}).

We now need to find a ground state of the D1-brane theory, and study fluctuations about this
ground state. First consider a single D1-brane. If in the Dirac-Born-Infeld action,
\be\label{DBI}
S_{D1}=-{1\over2\pi\ell_s^2}\int d\tau d\sigma e^{-\phi}\sqrt{-\det\left(\partial_\alpha
X^\mu\partial_\beta X^\nu G_{\mu\nu}+2\pi\ell_s^2 F_{\alpha\beta}
\right)},
\ee
we first put $F_{\alpha\beta}$ to zero, then the $x^+$ dependence cancels
between the inverse string coupling and the determinant of the metric.
So a simple classical solution is
\bea
x^1 &= &{1\over r}\sigma,\\
x^+ &=&{1\over r} {\tau \over \sqrt2},\\
x^- &=& {1\over r}{\tau \over \sqrt2},\\
x^i &=& 0,\ \ \  i=2,\ldots,8.
\eea
If we now choose the gauge
\bea
x^1 &= &{1\over r}\sigma,\nonumber\\
x^+ &=& {1\over r}{\tau \over \sqrt2},\label{gaugexplus}
\eea
and define a new coordinate $y$ by
\be
x^- ={1\over r} {\tau \over \sqrt2} + \sqrt2 y,
\ee
then, ignoring a total derivative,~\C{DBI}\ can be expanded to give
\bea
S_{D1}&=&{1\over2\pi\ell_s^2}\int d\tau d\sigma\left(
-{1\over r^2}+\half\left[(\partial_\tau y)^2+(\partial_\tau x^i)^2-(\partial_\sigma y)^2-(\partial_\sigma x^i)^2\right]\right.\nonumber\\
&&\ \ \ \ \ \ \ \ \ \ \ \ \ \ \ \ \ \ \ \ \ \left.
+2\pi^2\ell_s^4\,\exp\left(-{\sqrt2\epsilon Q\tau\over r}\right) F_{\tau\sigma}^2
+\ldots\right),\label{YMmatrix}
\eea
with
\be
\sigma\sim\sigma+ 2\pi\ell_s.
\ee

This agrees with the $N=1$ case of \eq{sym} after rescaling the fields.
Note that the coordinate $y$ appears on the same footing as the $x^i$ ($i=2,\ldots,8$), so
it plays the role that $x^i$ used to play before we made the compactification
space-like.

{}For $N$ D1-branes, the world-volume theory is given by~\eq{sym}.

\subsection{Regime of validity of the Matrix string description}
\label{decargument}
The modes \eq{basissoln}
of a scalar field in the lightlike linear dilaton background
(\ref{Mink}, \ref{lld}) take the following form in terms of the
new coordinates \eq{basistransf}:
\be\label{Tnew}
T(x^+,x^-,\vec x)=e^{-\epsilon Qx^+}\exp\left\{-i(\epsilon E^-
+{p^+\over2\epsilon}-k_1)x^+-i{p^+\over\e}x^-+i(k_1-{p^+\over\e})x^1
+i\sum_{j=2}^8k_jx^j\right\}.
\ee
The identification \eq{identify}, or equivalently
\eq{identifybis}, implies the momentum quantization condition
\be
p^+=\epsilon k_1-{2\pi n\over R}.
\ee
In DLCQ, we focus on a sector with given $N$ and study
fluctuations that stay within that sector. Such fluctuations
have $n=0$, so that
\be\label{quantmom}
p^+=\epsilon k_1.
\ee
Using \eq{quantmom}, the mass shell condition \eq{eeeen} for
non-negative $m^2$ implies
\be
|k_1|\leq 2\e |E^-|.
\ee
As a consequence, \eq{Tnew} shows that the energy and momentum in
the new coordinate system $(x^+,x^-,x^i)$ are at most of order
\be\label{STenergy}
\e E^-.
\ee
Taking into account the identifications \eq{gaugexplus},
we conclude that the world-sheet energy and momentum appearing in the action \eq{YMmatrix}
are at most of order
\be\label{energy}
E_{typical}\sim{\e E^-\over r}\sim {E^-\ell_s\over R}.
\ee

The effective time-dependent string length $\ell_s^{\rm eff}$ can be read from the metric~\C{metricD1},
\be
\ell_s^{\rm eff} = {\ell_s e^{-\e Qx^+/2}  \over \sqrt{r}}.
\ee
The condition for open string oscillators to decouple is given by
\be \label{opengone}
\e E^- \ell_s^{\rm eff}  = {\sqrt{2\pi \e \ell_s^3 \over R} E^- e^{-\e Qx^+/2}}\ll 1,
\ee
which is satisfied in the $\epsilon\to 0$ limit.

We must also check that gravity decouples from the Matrix description.
The effective ten-dimensional Newton ``constant'' can be determined from the metric~\C{metricD1}\ and
dilaton~\C{dilneg},
\be
G_N^{\rm eff} \sim g_s^2 (\ell_s^{\rm eff})^8 = {\ell_s^8e^{-2\e Qx^+}\over r^2}= {4\pi^2\ell_s^{10} e^{-2\e Qx^+}\over \e^2 R^2}.
\ee
Taking into account the fact that the energies of fluctuations are given by \eq{STenergy}, we see that
the fluctuations interact gravitationally with strength
\be\label{decoupclosed}
G_N (\e E^-)^8\sim {4\pi^2\ell_s^{10}\e^6 (E^-)^8e^{-2\e Qx^+}\over R^2},
\ee
so closed strings also decouple for $\epsilon\to0$.

So complete decoupling of closed and massive open strings can be achieved by strictly setting
\be\label{limitzero}
\e=0.
\ee
This strict limit makes perfect sense from the SYM point of view, since its coupling
\be\label{gym}
g_{YM}={1\over g_s\ell_s}={1\over \ell_s}\exp\left({\sqrt2\pi\ell_sQ\tau\over R}\right)
\ee
and the typical energies \eq{energy} are $\e$-independent. In other words, by using the limit \eq{limitzero},
we reach the remarkable conclusion that our Matrix description is valid all the way to the singularity, which opens up
the perspective of using it to study the fate of the singularity.

Now that we have argued that the Matrix description is a complete description of the physics of the singularity, we
should ask
whether the description is weakly coupled. The relevant dimensionless parameter is the ratio of the SYM coupling to the
typical energy of processes we are interested in. Using \eq{gym}, \eq{energy}, \eq{pplus} and
\be\label{pplus}
p^+=2\pi N/R,
\ee
we see that this parameter equals
\be\label{param}
{g_{YM}\over {E_{typical}}}\sim{N\exp\left({\sqrt2\pi\ell_sQ\tau\over R}\right)\over p^+E^-\ell_s^2}.
\ee
Thus for any fixed finite $N$ the Matrix description is strongly coupled for late times and weakly coupled for early
times.
Note, however, that in DLCQ, we eventually want to take the decompactification limit $N\to\infty$, $R\to\infty$ with
fixed $p^+=2\pi N/R$. This is corresponds to the limit where we take
$N\to \infty$ holding $g^2_{YM}$ fixed while considering energies of order $1/N$.
In a strict $N\to\infty$ limit, which undoes the DLCQ by decompactifying the lightlike circle, one finds a
strong coupling behavior for all times. One could also imagine having $N$ depend on time and keeping an appropriate
combination of the parameter \eq{param} and $N$ almost fixed close to the singularity, thus effectively decompactifying
the DLCQ circle near the singularity while keeping the SYM theory weakly coupled near the singularity. For some
range of times, it might
also be useful to perform an analysis along the lines of~\cite{Itzhaki:1998dd}.

\subsection{Cosmological evolution and the emergence of space-time}

Time evolution from the big bang to late times corresponds beautifully
to renormalization group flow in the Yang-Mills theory. At the big bang, the Yang-Mills theory is
weakly coupled. Since the coupling has positive mass dimension, weak coupling corresponds to the UV
sector of the theory. As time evolves, the coupling increases and the Yang-Mills theory flows to the IR.

At early times, we have weakly coupled Yang-Mills theory. Since the coupling~\C{ymcoupling}\ is small, the
potential terms in~\C{sym}\ turn off as we approach the big bang. We are left with a theory of non-commuting
matrices. These appear to be the correct degrees of freedom near the singularity, replacing our conventional
notion of space-time.
At very late times, we recover light-cone quantized perturbative string field theory in the light-like
linear dilaton background, along the lines described in~\cite{Dijkgraaf:1997vv}.

In Matrix theory descriptions of flat space-time, supersymmetry plays a critical role.
Roughly speaking, diagonal matrix elements are interpreted as positions in space-time (they are
the positions of the D1-branes in their transverse space), while off-diagonal matrix elements
correspond to strings connecting the various D1-branes. When two well-separated clusters of D1-branes
are considered, for instance corresponding to two well-separated supergravitons, the off-diagonal
modes between the two clusters are heavy and can be integrated out, a
priori giving rise to an effective potential for the separation modulus of the two clusters. Supersymmetric
cancellations ensure that the effective potential vanishes and that the moduli space metric is
flat~\cite{Paban:1998ea}.
This is crucial for the space-time interpretation
of Matrix theory: if supergravitons interacted with a static rather than velocity-dependent potential, the
model would not describe gravity in flat space-time.

Indeed, a Matrix theory description of a non-supersymmetric flat string background with a closed string tachyon
was given in~\cite{Banks:1999tr}. This Matrix theory develops a potential that lifts the flat directions, a
pathology that was given the interpretation that the original non-supersymmetric space-time is not a solution
of non-perturbative string theory.

Our model, the light-like linear dilaton background of type IIA string theory, preserves 16 supersymmetries, so one
might have hoped to find a supersymmetric Matrix theory description. However, it turns out that supersymmetry is
spontaneously broken in any sector with non-zero light-cone momentum $p^+=2\pi N/R$. Since in our discrete light-cone
quantization we focus on a sector with a fixed non-zero value for $N$, we are bound to find a Matrix string
description in which supersymmetry is broken. This would have been disastrous if the breaking were explicit.

However, what we actually found is maximally supersymmetric two-dimensional Yang-Mills theory in flat space
with a boost identification. The boost identification breaks all the supersymmetry but only via boundary conditions
on the Milne circle: namely, the action of the boost transformation acts differently on fields of different spin
leading to a different quantization condition for bosons and fermions~\cite{Berkooz:2004re}. This is an effect that
becomes less relevant as the Milne circle grows in time. At late times, the potential becomes small as supersymmetry
is restored. It would be interesting to understand this potential in a quantitative way.

\subsection{Does time begin?}

The deepest question that we would hope to address with this formalism is whether the big bang should be thought
of as the beginning of time, or whether space-time exists prior to the singularity. From a space-time point of view,
it is not clear that it makes sense to continue the spacetime metric \eq{11metbis} beyond the singularity at
$\lambda=0$.

The same question has been addressed for the Milne orbifold as (two dimensions of) a space-time background
in perturbative string theory \cite{Horowitz:2002mw, Berkooz:2002je}. The conclusion was that $2\rightarrow2$
scattering amplitudes
across the singularity
diverge at tree-level because of large tree-level gravitational backreaction from the region close to the singularity
\cite{Berkooz:2002je}.

Our Matrix description a priori contains only the future quadrant of the Milne orbifold as the Matrix string
world-sheet.
One could try to include the other three quadrants of the Milne orbifold and see whether states can be propagated
across the singularity. Although there is now no gravitational backreaction, one can show along the lines
of~\cite{Berkooz:2002je} that large gauge backreaction
gives rise to UV enhanced IR divergences similar to the ones found in~\cite{Liu:2002kb}. These divergences are
in some sense milder than those encountered in the gravitational case, but they might still be problematic in our
low dimension field theory. However, it is not entirely
clear that these divergences are associated with the singularity.

It is also possible that the correct prescription involves selecting an initial condition at the big bang and
considering only the future quadrant of the Milne orbifold. There are two natural states to choose as an initial
vacuum: the conformal vacuum annihilated by positive frequency modes with respect to the conformal time $\eta$ given
in~\C{confmet}, and the adiabatic vacuum inherited from the underlying Minkowski space. For a more detailed
discussion, see~\cite{Birrell:1982ix, Berkooz:2002je}. At the big bang, we have a conformal field theory since
the Yang-Mills theory is free. The natural vacuum from the perspective of conformal field theory would be the
conformal vacuum which is $SL(2)$ invariant. However, the
adiabatic vacuum has the advantage of better high-energy behavior at the singularity, and is the vacuum state that
is usually used in string theory computations~\cite{Nekrasov:2002kf, Berkooz:2002je}.
At late times, where the perturbative string description is
good, it is natural to study states with reference to conformal time which is identified with $X^+$ in
space-time via~\C{gaugexplus}.

It is an interesting question to determine the precise observables in this Matrix model. For example, one
possibility could be to
take a natural initial Yang-Mills state and ask how it evolves into a collection of  excited strings in the
space-time that emerges at late times.

\section{Some Generalizations}
\subsection{A dual type IIA background}

There are many interesting ways to generalize the light-like linear dilaton solution. For
example, given the M-theory metric (\ref{11met}), we can compactify the $X^9$ direction
and interpret it as the M-theory circle. This gives rise to an alternate
type IIA description, with metric
\be\label{tildemet}
{ds}_{10}^2=e^{QX^+}[-(dX^0)^2+(dX^1)^2+\cdots+(dX^8)^2]+e^{-QX^+}(dY)^2
\ee
and light-like linear dilaton
\be
\phi= {QX^+\over2}.
\ee
In a coordinate basis, the metric (\ref{tildemet}) has non-vanishing
curvature components
\bea
R_{+i+i} &=& {Q^2\over 4} e^{QX^+}, \cr
R_{+y+y} &=&- {3  Q^2 \over 4} e^{-QX^+}.
\eea
Thus we find that the Ricci tensor has a non-vanishing component
\be
R_{++}=Q^2.
\ee
The non-vanishing Ricci tensor is supported by the dilaton, which in the
presence of the metric (\ref{tildemet}) makes a non-vanishing
contribution to Einstein's equations:
\be
\nabla_+\nabla_+\phi=-\Gamma^+_{++}\partial_+\phi=-{Q^2\over2}.
\ee
Like (\ref{11met}), the space-time (\ref{tildemet}) is singular as
$X^+\to -\infty$ because the metric components $\tilde g_{ii}$ go
to zero in that limit.

If we now compactify the $X^8$-direction, $X^8\sim X^8+2\pi  R^8$, and T-dualize, we find a type IIB solution
with metric
\be
{ds}_{10}^2=e^{QX^+}[-(dX^0)^2+(dX^1)^2+\cdots+(dX^7)^2]+e^{-QX^+}[(dX^8)^2+(dY)^2]
\ee
and constant dilaton,
\be
\phi = \log\left({\ell_s\over R^8}\right)
\ee
with $X^8\sim X^8+2\pi\ell_s^2/R^8$.

\subsection{The light-like linear dilaton in type IIB}

We could also directly consider the light-like linear dilaton in type IIB string theory.
The new ingredient in type IIB that we should consider
as $g_s \rr \infty$ is S-duality. The Einstein frame metric~\C{einstein}\ is
invariant under S-duality, but we obtain a new string frame metric
valid in the strong coupling regime. We can view the resulting S-dual
description as either string theory with a time-independent string length
$\ell_s$ but with a coupling and metric
\be
\widetilde{g_s} = e^{Q X^+}, \qquad ds^2 =  e^{Q X^+}\left\{
-2 dX^+ dX^- + \sum_{i} (dX^i)^2 \right\},
\ee
or as string theory with a coupling and metric
\be
\widetilde{g_s} = e^{Q X^+}, \qquad ds^2 = \left\{
-2 dX^+ dX^- + \sum_{i} (dX^i)^2 \right\},
\ee
and a time-dependent string length $\widetilde{\ell_s}^2 = e^{-Q X^+}
\ell_s^2. $ As we approach the singularity at $X^+ \rr -\infty$, the
effective string tension goes to zero. Once again, we see that the
resolution of the singularity requires physics beyond the
$\widetilde{\ell_s}$ expansion.

It might appear that because the string coupling is small near the singularity,
we should have a good perturbative string description. If this were the
case, we might hope to resolve the cosmological singularity in string
perturbation theory. However, this is not the case:
graviton perturbation  theory,  which is controlled by the (duality invariant)
effective Newton constant, still breaks down near the singularity. We
therefore expect a new
description involving new degrees of freedom in the strong coupling
regime. Such a description should follow from type IIB Matrix string theory~\cite{Sethi:1997sw,
Banks:1996my}.

\subsection{The light-like linear  dilaton from little string theory}

Finally, it is worth mentioning that the light-like linear dilaton can be obtained as an unusual Penrose
limit of the near
horizon geometry of type II NS5-branes studied in~\cite{NS5}.\footnote{We would like to
thank Daniel Robbins for collaboration on the material in this subsection.} This geometry is described by
\be
ds^2 = N\ell_s^2\left[-d\tilde t^2+\frac{dr^2}{r^2}+d\mO_3^2\right]+dy_5^2, \qquad
e^{2\Phi}=\frac{N\ell_s^2g_s^2}{r^2}.
\ee
The conventional time coordinate $t$ has been rescaled, $t=\sqrt N\ell_s\tilde t$,  to uniformize the factors of $N$ appearing in the metric.  There is also an NS three-form field strength $H_3$
which is $N$ times the volume form of the three-sphere, but this will vanish
when we take the limit.

We are interested in boosting along a radial rather than angular null geodesic in this
space.  To do this we switch coordinates $(\tilde t,r)\rightarrow(u,v)$ by
$\tilde t=u-v$, $r=\sqrt N\ell_se^u$.  This gives
\be
ds^2 = N\ell_s^2\left[2dudv-dv^2+d\theta^2+\cos^2\theta d\psi^2+
\sin^2\theta d\phi^2\right]+dy_5^2, \quad e^{2\Phi}=g_s^2e^{-2u}.
\ee
Finally, to take the limit \cite{NS5}, we rescale $v\rightarrow v/N$,
$\theta\rightarrow\theta/\sqrt N$, $\psi\rightarrow\psi/\sqrt N$, and take $N\rightarrow\infty$.
This sends $H_3$ to zero since
\be H_3\sim N\sin\theta d\theta d\psi d\phi,\ee
which goes like $N^{-1/2}$. We are left with the metric,
\be
ds^2 = \ell_s^2\left[2dudv+dx_3^2 \right]+dy_5^2, \qquad \Phi=\Phi_0-u,
\ee
which describes the light-like linear dilaton in flat space-time.

If we were now to study perturbative states in light-cone string theory on this
background, we would like to understand to which states these correspond in
the original coordinates.  After the limit, the light cone energy and momentum
are given by
\bea
2p^- &=& -i\frac{\p}{\p u} = -i\left(\frac{\p}{\p\tilde t}+r\frac{\p}{\p r}
\right) = -i\left(\ell_s\sqrt N\frac{\p}{\p t}+r\frac{\p}{\p r}\right)\non\\
2p^+ &=& -\frac{i}{N}\frac{\p}{\p v} = \frac{i}{N}\frac{\p}{\p\tilde t} =
\frac{i}{\sqrt N}\ell_s\frac{\p}{\p t}.
\eea
The states of interest to us are those with finite $p^-$ and $p^+$.  Mapping these states back to the original
variables, we see that these states comprise a
certain sector of high-energy states in little string theory~\cite{Seiberg:1997zk}\ with energies of order
$\sqrt{N}$.

The main problem with this approach for obtaining a holographic description of the null dilaton is that little
is still known about little string theory beyond its bulk definition. However, many other geometries give rise to
the light-like linear dilaton via similar limits, and perhaps one of those geometries will provide a more
tractable holographic dual in the spirit of the AdS/CFT correspondence.

\section*{Acknowledgements}

The work of B.~C. is supported by Stichting FOM.
The work of S.~S. is supported in
part by NSF CAREER Grant No. PHY-0094328, and by the Alfred P. Sloan
Foundation.

B.~C. and S.~S. would like to thank the Aspen Center for Physics
for hospitality during the early stages of this work.
B.C.\ thanks the organizers  of the
IPAM ``Conformal Field Theory 2nd Reunion Conference'' at Lake Arrowhead,
where part of this work was carried out.
S.~S. would also like to thank the
organizers of the 2005 Amsterdam String Theory Workshop, where this work was completed.

\providecommand{\href}[2]{#2}\begingroup\raggedright\endgroup



\end{document}